\documentclass{article}
\usepackage{ijcai21}

\usepackage{times}
\usepackage{soul}
\usepackage{url}
\usepackage[hidelinks]{hyperref}
\usepackage[utf8]{inputenc}
\usepackage[small]{caption}
\usepackage{graphicx}
\usepackage{amsmath}
\usepackage{amsthm}
\usepackage{booktabs}
\usepackage{algorithm}
\usepackage{algorithmic}
\usepackage{multirow}
\title{Beyond a binary of (non)racist tweets: A four-dimensional categorical detection and analysis of racist and xenophobic opinions on Twitter in early Covid-19}

\author{
    Xin Pei$^1$ and Deval Mehta$^2$\\
    \affiliations
    $^1$School of International Communications, University of Nottingham, Ningbo, China\\
    $^2$Independent Researcher
    \emails
    PEIX0001@gmail.com, deval.mehta1092@gmail.com
}

\begin{document}

\maketitle

\begin{abstract}
Transcending the binary categorization of racist and xenophobic texts, this research takes cues from social science theories to develop a four-dimensional category for racism and xenophobia detection, namely stigmatization, offensiveness, blame, and exclusion. With the aid of deep learning techniques, this categorical detection enables insights into the nuances of emergent topics reflected in racist and xenophobic expression on Twitter. Moreover, a stage wise analysis is applied to capture the dynamic changes of the topics across the stages of early development of Covid-19 from a domestic epidemic to an international public health emergency, and later to a global pandemic. The main contributions of this research include, first the methodological advancement. By bridging the state-of-the-art computational methods with social science perspective, this research provides a meaningful approach for future research to gain insight into the underlying subtlety of racist and xenophobic discussion on digital platforms. Second, by enabling a more accurate comprehension and even prediction of public opinions and actions, this research paves the way for the enactment of effective intervention policies to combat racist crimes and social exclusion under Covid-19. 
\end{abstract}

\section{Introduction}

The rise of racism and xenophobia has become a remarkable social phenomenon stemming from Covid-19 as a global pandemic. Especially, attention has been increasingly drawn to the Covid-19 related racism and xenophobia which has manifested a more infectious nature and harmful consequences compared to the virus itself \cite{wang2021m}. According to BBC report, throughout 2020, anti-Asian hate crimes increased by nearly one hundred and fifty percent, and there were around three thousand eight hundred anti-Asian racist incidents. Therefore, it has become urgent to comprehend public opinions regarding racism and xenophobia for the enactment of effective intervention policies preventing the evolvement of racist hate crimes and social exclusion under Covid-19. Social media as a critical public sphere for opinion expression provides platform for big social data analytics to understand and capture the dynamics of racist and xenophobic discourse alongside the development of Covid-19. 

This research agenda has drawn attention from an increasing body of studies which have regarded Covid-19 as a social media infodemic \cite{cinelli2020covid}, \cite{gencoglu2020causal}, \cite{trajkova2020exploring}, \cite{li2020analyzing}, \cite{guo2021covid}. The work in \cite{schild2020go} made an early and probably the first attempt to analyse the emergence of Sinophobic behaviour on Twitter and Reddit platforms. Soon after \cite{ziems2020racism} studied the role of counter hate speech in facilitating the spread of hate and racism against the Chinese and Asian community. The authors in \cite{vishwamitra2020analyzing} attempted to study the effect of hate speech on Twitter targeted on specific groups such as the older community and Asian community in general. The work in \cite{pei2020coronavirus} demonstrated the dynamic changes in the sentiments along with the major racist and xenophobic hashtags discussed across the early time period of Covid-19. The authors in \cite{masud2020hate} explored the user behavior which triggers the hate speech on Twitter and later how it diffuses via retweets across the network. All these methods have used highly advanced computational techniques and state-of-the-art language models for extracting insights from the data mined from Twitter and other platforms.

While focusing on technical advancement, many studies tend to neglect the foundation for accurate data detection and analysis – that is how to define racism and xenophobia. Especially, the computational techniques and models tend to apply a binary definition (either racist or non-racist) to categorise the linguistic features of the texts, with limited attention paid to the nuances of racist and xenophobic behaviours. However, understanding the nuances is critical for mapping the comprehensive picture of the development of racist and xenophobic discourse alongside the evolvement of Covid-19 – whether and how the expression of racism and xenophobia may change the topics across time. More importantly, capturing these changes reflected in the online public sphere will enable a more accurate comprehension and even prediction of public opinions and actions regarding racism and xenophobia in the offline world. 

 Reaching this goal demands a combination of computational methods and social science perspectives, which becomes the focus of this research. With the aid of BERT (Bi-directional Encoder Representations from Transformers) \cite{devlin2018bert} and topic modelling \cite{blei2003latent}, The main contribution of this research lies in two aspects: 
\begin{enumerate}
    \item Development of a four-dimensional categorization of racist and xenophobic texts into stigmatization, offensiveness, blame, and exclusion; 
    \item Performing a stage wise analysis of the categorized racist and xenophobic data to capture the dynamic changes amongst the discussion across the development of COVID-19.
\end{enumerate}

Especially, this research situates the examination in Twitter, the most influential platform for political online discussion. And we focus on the most turbulent early phase of Covid-19 (Jan to Apr 2020) where the unexpected and constant global expansion of virus kept on changing people's perception of this public health crisis and how it is related to race and nationality. To specify, this research divides the early phase into three stages based on the changing definitions of Covid-19 made by World Health Organization (WHO) - (1) 1st to 31st Jan 2020 as a domestic epidemic referred to as stage 1 (S1); (2) 1st Feb to 11th Mar 2020 as an International Public Health Emergency (after the announcement made by WHO on 1st Feb) referred to as stage 2 (S2); (3) 12th Mar to 30th Apr 2020 as a global pandemic (based on the new definition given by WHO on 11th Mar) referred to as stage 3 (S3).

The rest of the paper is organized as follows. In section 2, we outline the dataset mined from Twitter. Section 3 deals with two parts - firstly, it presents the data and method employed for category-based racism and xenophobia detection. Secondly, it details the topic modelling employed for extracting topics from the categorized data. In section 4, we discuss the findings of the overall process with the focus on topics emerging amongst the different racism and xenophobia categories across the early development of Covid-19. Finally, we conclude this paper in section 5.

\section{Dataset}
Dataset of this research is comprised of 247,153 tweets extracted through Tweepy API\footnote{https://www.tweepy.org/} from the eighteen most circulated racist and xenophobic hashtags related to Covid-19 from 1st January to 30th April in the year of 2020. The list of selected hashtags is as follows - \texttt{\#}chinavirus, \texttt{\#}chinesevirus, \texttt{\#}boycottchina, \texttt{\#}ccpvirus, \texttt{\#}chinaflu, \texttt{\#}china\_is\_terrorist, \texttt{\#}chinaliedandpeopledied, \texttt{\#}chinaliedpeopledied, \texttt{\#}chinalies, \texttt{\#}chinamustpay, \texttt{\#}chinapneumonia, \texttt{\#}chinazi, \texttt{\#}chinesebioterrorism, \texttt{\#}chinesepneumonia, \texttt{\#}chinesevirus19, \texttt{\#}chinesewuhanvirus, \texttt{\#}viruschina, and  \texttt{\#}wuflu. The extracted tweets from the above hashtags are further divided into three stages that define the early development of Covid-19 as mentioned earlier.  

\begin{table*}
\centering
\resizebox{\textwidth}{!}{\begin{tabular}{l|p{80mm}p{80mm}}
\hline
Category & Definition & Example \\
\hline
Stigmatization & Confirming negative stereotypes for conveying \newline a devalued social identity within a particular context\cite{miller2001theoretical} & “For all the \#ChinaVirus jumped from a bat at the wet market”\\
\hline
Offensiveness & Attacking a particular social group \newline through aggressive and abusive language \newline \cite{jeshion2013expressivism} & “Real misogyny in communist China. \#chinazi \#China\_is\_terrorist \#China\_is\_terrorists \#FuckTheCCP” \\
\hline
Blame & Attributing the responsibility for the \newline negative consequences of the crisis to one social group \newline \cite{coombs2000empirical} & “These Chinese are absolutely disgusting. They spread the \#ChineseVirus. Their lies created a pandemic \#ChinaMustPay” \\
\hline
Exclusion & the process of othering to draw a clear boundary \newline between in-group and out-group members \newline \cite{bailey2005racialised} & “China deserves to be isolated by all means forever. SARS was also initiated in China, 2003 by eating anything \& everything \#BoycottChina” \\
\hline
\end{tabular}}
\caption{Definition and example of categorization of racist and xenophobic behaviors.}
\label{tab:def_racism}
\end{table*}

\section{Method}
\subsection{Category-based racism and xenophobia detection}
Beyond a binary categorization of racism and xenophobia, this research applies the perspective of social science to categorizing racism and xenophobia into four dimensions as demonstrated in Table \ref{tab:def_racism}. This basically translates into a problem of five class classification of text data, where four classes represent the racism and xenophobia categories and fifth class corresponds to the category of non-racist and non-xenophobic.

\subsubsection{Annotated dataset}
For this purpose, we annotate a dataset of 6000 tweets. These tweets were randomly selected from all hashtags across the three development stages, and annotated by four research assistants with inter-coder reliability reaching above \textbf{70\%}. The annotation followed a coding method with 0 representing stigmatization, 1 for offensiveness, 2 for blame, and 3 for exclusion in alignment with the linguistic features of the tweets. The non-marked tweets were regarded as non-racist and non-xenophobic and represented class category 4. We limit the annotation for each tweet to only one label which aligns to the strongest category. The distribution of 6000 tweets amongst the five classes is as follows - 1318 stigmatization, 1172 offensive, 1045 blame, 1136 exclusion, and 1329 non-racist and non-xenophobic.

We view the task of classification of the above-mentioned categories as a supervised learning problem and target developing machine learning and deep learning techniques for the same. We firstly pre-process the input data text by removing punctuation and URLs from a text sample and converting it to lower case before providing it to train our models. We split the data into random train and test splits with 90:10 ratio for training and evaluating the performance of our models respectively.

\subsubsection{BERT}
\begin{figure}
\centering
\includegraphics[trim=0.1cm 0.2cm 0.1cm 0.3cm,clip,width=1.0\linewidth,keepaspectratio]{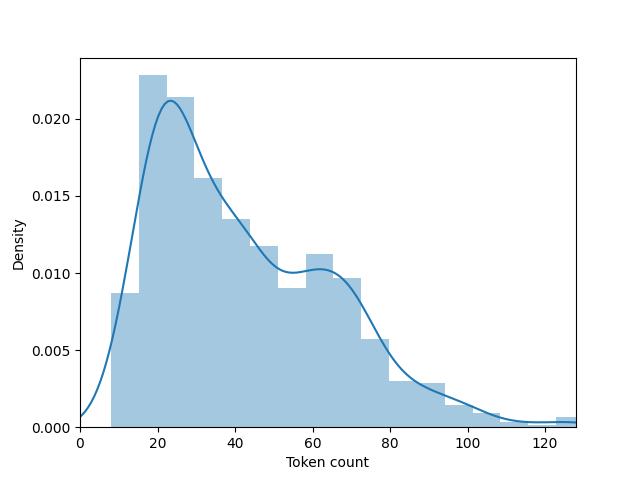}
\caption{Density distribution of token lengths of the tweets in our dataset.}
\label{fig_token_len}
\end{figure}

Recently, word language models such as Bi-directional Encoder Representations from Transformers (BERT) \cite{devlin2018bert} have become extremely popular due to their state-of-the-art performance on natural language processing tasks. Due to the nature of bi-directional training of BERT, it can learn the word representations from unlabelled text data powerfully and enables it to have a better performance compared to the other machine learning and deep learning techniques \cite{devlin2018bert}. The common approach for adopting BERT for a specific task on a smaller dataset is to fine-tune a pre-trained BERT model which has already learnt the deep context-dependent representations. We select the “bert-base-uncased” model which comprises of 12 layers, 12 self-attention heads, a hidden size of 768 totalling 110M parameters. We fine-tune the BERT model with a categorical cross-entropy loss for the five categories. The various hyperparameters used for fine-tuning the BERT model are selected as recommended from the paper \cite{devlin2018bert}. We use the AdamW optimizer with the standard learning rate of 2e-5, a batch size of 16, and train it for 5 epochs. For selecting the maximum length of the sequences, we tokenize the whole dataset using Bert tokenizer and check the distribution of the token lengths. We notice that the minimum value of token length is 8, maximum is 130, median is 37 and mean is ~42. Based on the density distribution shown in Fig.\ref{fig_token_len}, we experiment with two values of sequence length – 64 and 128 and find that the sequence length of 64 provides a better performance.

As additional baselines, we also train two more techniques. Long Short Term Memory Networks (LSTMs) \cite{hochreiter1997long} have been very popular with text data as they can learn the dependencies of various words in the context of a text. Also, machine learning algorithms such as Support Vector Machine (SVMs) \cite{hearst1998support} have been used previously by researchers for text classification tasks. We adopt the same data pre-processing and implementation technique as mentioned earlier and train the SVM with grid search, a 5-layer LSTM (using the pre-trained Glove \cite{pennington2014glove} embeddings) and BERT model for the category detection of the racist and xenophobic tweets.

For evaluating the machine learning and deep learning approaches on our test dataset, we use the metrics of average accuracy and weighted f1-score for the five categories. The performance of the model is shown in Table \ref{tab:performance}. It can be seen from Table \ref{tab:performance} that the fine-tuned BERT model performs the best compared to SVM and LSTM in terms of both accuracy and f1 score. Thus, we employ this fine-tuned BERT model for categorizing all the tweets from the remaining dataset. Having employed BERT on the remaining dataset, we get a refined dataset of the four categories of tweets spreaded across the three stages as shown in Table \ref{tab:distribution}.

\begin{table}
\centering
\begin{tabular}{lcc}
\hline
Technique & Accuracy(\%) & F1-score \\
\hline
SVM & 69 & 0.66 \\
LSTM & 74 & 0.72 \\
BERT & \textbf{86} & \textbf{0.81} \\
\hline
\end{tabular}
\caption{Performance of different models on the manually annotated test dataset.}
\label{tab:performance}
\end{table}

\begin{table}
\centering
\begin{tabular}{lcccc}
\hline
Category & Total & S1 & S2 & S3 \\
\hline
Stigmatization & 116584 & 3723 & 5687 & 107174 \\
Offensiveness & 10503 & 1722 & 1808 & 6973\\
Blame & 39765 & 31 & 777 & 38957\\
Exclusion & 10293 & 872 & 1341 & 8080 \\
\hline
\end{tabular}
\caption{Distribution of tweets amongst the four categories across the three stages.}
\label{tab:distribution}
\end{table}

\begin{table*}
\centering
\resizebox{\textwidth}{!}{
\begin{tabular}{p{5mm}|l|cccccccccc}
\hline
\multirow{5}{*}{S1} & T1.\textbf{Virus} & virus & spread & country & travel & year & control & \textit{chinese} & ban & corona & show\\

& T2.\textbf{China/Chinese} & \textit{chinese} & virus & deadly & \textit{china} & situation & mask & stop & animal & source & eat\\

& T3.\textbf{Infection} & people & case & health & infect & confirm & death & sar & number & report & market\\

& T4.\textbf{Outbreak} & \textit{china} & coronavirus & wuhan & outbreak & city & hospital & news & patient & put & state\\

& T5.\textbf{Travel} & world & \textit{china} & government & make & people & time & day & bad & flight & start\\
\hline
\multirow{5}{*}{S2} & T1.\textbf{Emergency} & virus & spread & day & year & corona & show & emergency & food & kit & supply\\

& T2.\textbf{Globe} & \textit{china} & world & time & country & report & death & global & health & travel & confirm\\

& T3.\textbf{Infection} & people & case & call & ncov & infect & kill & pack & state & flu & number\\

& T4.\textbf{China} & \textit{china} & coronavirus & wuhan & outbreak & quarantine & stop & find & man & dead & thing\\

& T5.\textbf{Chinese} & \textit{chinese} & make & mask & government & news & good & work & citizen & start & respirator\\
\hline
\multirow{5}{*}{S3} & T1.\textbf{Government} & \textit{china} & world & spread & country & lie & pay & communist & government & ccp & make\\

& T2.\textbf{?} & time & make & \textit{india} & good & give & work & day & back & fight & buy\\

& T3.\textbf{China} & \textit{china} & coronavirus & case & death & covid & country & economy & war & number & wuhan\\

& T4.\textbf{Chinese} & \textit{chinese} & virus & people & call & stop & racist & start & die & blame & corona\\

& T5.\textbf{US} & \textit{american} & trump & \textit{state} & medium & president & \textit{america} & news & great & propaganda & show\\
\hline

\end{tabular}
}
\caption{Extracted topics and their corresponding keywords for the category of stigmatization spread across the three stages S1, S2, and S3.}
\label{tab:topics_stigma}
\end{table*}

\begin{table*}
\centering
\resizebox{\textwidth}{!}{\begin{tabular}{p{5mm}|l|cccccccccc}
\hline
\multirow{5}{*}{S1} & T1.\textbf{?} & country & ccp & citizen & virus & arrest & live & system & security & foreign & understand\\

& T2.\textbf{Government} & people & government & democracy & support & life & year & regime & \textit{uyghur} & camp & give\\

& T3.\textbf{?} & \textit{china} & world & spread & stop & communist & happen & \textit{taiwan} & wuhan & govt & ban\\

& T4.\textbf{Muslim} & \textit{chinese} & make & muslim & good & kill & police & terrorist & bad & party & lie\\

& T5.\textbf{Human right} & world & freedom & \textit{hong\_kong} & human & human\_right & time & free & stand & \textit{hk} & fight\\
\hline
\multirow{5}{*}{S2} & T1.\textbf{Freedom} & world & stop & freedom & truth & spread & good & free & \textit{hk} & speech & life\\

& T2.\textbf{Ccp} & \textit{china} & \textit{chinese} & ccp & virus & happen & \textit{wuhan} & evil & communist & time & \textit{uyghur}\\

& T3.\textbf{People} & people & make & kill & lie & ppl & trust & camp & police & thing & man\\

& T4.\textbf{China} & \textit{china} & country & regime & pay & money & outbreak & start & work & force & control\\

& T5.\textbf{Human right} & government & citizen & human & fight & support & hong\_kong & taiwan & give & democracy & death\\
\hline
\multirow{5}{*}{S3} & T1.\textbf{Death} & world & people & pay & lie & kill & truth & fight & life & die & humanity\\

& T2.\textbf{Government} & time & call & government & \textit{india} & communist & pandemic & give & global & send & real\\

& T3.\textbf{Virus} & \textit{chinese} & virus & spread & \textit{wuhan} & corona & product & buy & control & big & day\\

& T4.\textbf{China} & \textit{china} & country & make & ccp & stop & good & coronavirus & human & trust & support\\

& T5.\textbf{World} & \textit{china} & world & war & case & start & covid & economy & death & state & \textit{italy}\\
\hline
\end{tabular}}
\caption{Extracted topics and their corresponding keywords for the category of offensiveness spread across the three stages S1, S2, and S3.}
\label{tab:topics_offense}
\end{table*}

\begin{table*}
\centering
\resizebox{\textwidth}{!}{\begin{tabular}{p{5mm}|l|cccccccccc}
\hline
\multirow{5}{*}{S1} & T1.\textbf{Lie} & lie & spread & virus & autocracy & deceit & imagine & true & horrible & infect & country\\

& T2.\textbf{Death} & \textit{china} & dead & die & day & order & monstrosity & true & thing & kong & high\\

& T3.\textbf{Safety} & coronavirus & move & lot & cvirus & epicenter & safety & march & careful & knowingly & health\\

& T4.\textbf{Time} & \textit{wuhan} & lunar\_new & sick & year & time & absolutely & medium & mutate & emperor & truth\\

& T5.\textbf{Infection} & people & \textit{chinese} & make & online & pandemic & catch & number & infect & community & official\\
\hline
\multirow{5}{*}{S2} & T1.\textbf{Government} & lie & \textit{chinese} & coronavirus & government & \textit{wuhan} & cover & day & body & thing & care\\

& T2.\textbf{Spread} & world & country & spread & happen & trust & kill & threat & steal & dead & face\\

& T3.\textbf{China} & \textit{china} & truth & bad & free & money & communist & case & find & start & move\\

& T4.\textbf{Virus} & virus & stop & make & control & good & \textit{china} & fight & live & report & human\\

& T5.\textbf{Death} & people & time & number & die & real & life & entire & back & citizen & death\\
\hline
\multirow{5}{*}{S3} & T1.\textbf{World} & world & \textit{china} & country & pay & pandemic & kill & global & economy & war\\

& T2.\textbf{?} & people & stop & human & \textit{american} & eat & put & president & market & happen & live\\

& T3.\textbf{Lie} & \textit{china} & lie & coronavirus & \textit{wuhan} & blame & die & case & cover & truth & number\\

& T4.\textbf{?} & make & time & \textit{china} & good & start & buy & trust & back & thing & country\\

& T5.\textbf{Government} & \textit{chinese} & virus & \textit{china} & government & call & communist & ccp & covid & spread & hold\\
\hline

\end{tabular}}
\caption{Extracted topics and their corresponding keywords for the category of blame spread across the three stages S1, S2, and S3.}
\label{tab:topics_blame}
\end{table*}

\begin{table*}
\centering
\resizebox{\textwidth}{!}{\begin{tabular}{p{5mm}|l|cccccccccc}
\hline
\multirow{5}{*}{S1} & T1.\textbf{Government} & support & gov & join & people & evil & time & stand & sanction & government & money\\

& T2.\textbf{Human right} & product & world & stop & human\_right & freedom & tag & good & challenge & ppl & economic\_infiltration\\

& T3.\textbf{Boycott} & \textit{china} & \textit{hong\_kong} & fight & regime & boycott & show & international & control & trust & communist\\

& T4.\textbf{Trade} & make & buy & ccp & day & thing & friend & \textit{taiwan} & \textit{japan} & hope & today\\

& T5.\textbf{Virus} & country & \textit{chinese} & people & spread & year & human & animal & protect & virus & eat\\
\hline
\multirow{5}{*}{S2} & T1.\textbf{Nation} & people & \textit{chinese} & animal & happen & government & initiative & nation & show & economy & law\\

& T2.\textbf{Virus} & virus & control & truth & support & live & kill & boycott & start & stand & cover\\

& T3.\textbf{Threat} & \textit{china} & time & lie & threat & company & trust & big & entire & spy & \textit{wuhan}\\

& T4.\textbf{Human right} & world & country & freedom & spread & human\_right & economic & thing & evil & steal & raise\\

& T5.\textbf{Trade} & make & product & stop & buy & day & \textit{china} & good & ccp & challenge & coronavirus\\
\hline
\multirow{5}{*}{S3} & T1.\textbf{Virus} & \textit{china} & virus & world & pay & spread & ccp & covid & corona & market & call\\

& T2.\textbf{Pandemic} & world & \textit{china} & company & communist & coronavirus & pandemic & global & nation & trust & war\\

& T3.\textbf{Trade} & \textit{chinese} & make & product & buy & boycott & stop & good & \textit{India} & economy & \textit{Indian}\\

& T4.\textbf{Human right} & people & lie & government & human & life & back & animal & kill & eat & bring\\

& T5.\textbf{China} & \textit{china} & country & time & start & business & give & thing & app & sell & money\\
\hline

\end{tabular}}
\caption{Extracted topics and their corresponding keywords for the category of exclusion spread across the three stages S1, S2, and S3.}
\label{tab:topics_exclude}
\end{table*}

\subsection{Topic modelling}

Topic modelling is one of the most extensively used methods in natural language processing for finding relationships across text documents, topic discovery and clustering, and extracting semantic meaning from a corpus of unstructured data \cite{jelodar2019latent}. Many techniques have been developed by researchers such as Latent Semantic Analysis (LSA) \cite{deerwester1990indexing}, Probabilistic Latent Semantic Analysis (pLSA) \cite{hofmann1999probabilistic} for extracting semantic topic clusters from the corpus of data. In the last decade, Latent Dirichlet Allocation (LDA) \cite{blei2003latent} has become a successful and standard technique for inferring topic clusters from texts for various applications such as opinion mining \cite{zhai2011constrained}, social medial analysis \cite{cohen2013classifying}, event detection \cite{lin2010pet} and consequently there have also been various developed variants of LDA \cite{blei2010supervised} and \cite{blei2003hierarchical}.

For our research, we adopt the baseline LDA model with Variational Bayes sampling from Gensim\footnote{https://pypi.org/project/gensim/} and the LDA Mallet model \cite{mccallum2002mallet} with Gibbs sampling for extracting the topic clusters from the text data. Before passing the corpus of data to the LDA models, we perform data pre-processing and cleaning which include the following steps. Firstly, we remove any new line characters, punctuations, URLs, mentions and hashtags. Later we tokenize the texts in the corpus and also remove any stopwords using the Gensim utility of pre-processing and stopwords defined in the NLTK\footnote{https://pypi.org/project/nltk/} corpus. Finally, we make bigrams and lemmatize the words in the text.

After employing the above pre-processing for our corpus, we employ topic modelling using LDA from Gensim and LDA Mallet. We perform experiments by varying the number of topics from 5 to 25 at an interval of 5 and checking the corresponding coherence score of the models as was done in \cite{fang2016exploring}.  We train the models for 1000 iterations with varying number of topics, optimizing the hyperparameters every 10 passes after each 100 pass period. We set the values of $\alpha$, $\beta$ which control the distribution of topics and the vocabulary words amongst the topics to the default settings of 1 divided by the number of topics. We notice from our experiments that LDA Mallet has a higher coherence score (0.60-0.65) compared to the LDA model from Gensim (0.49-0.55) and thus we select LDA Mallet model for the task of topic modelling on our corpus of data.

The above strategy is employed for each racist and xenophobic category and for every stage individually. We find the highest coherence score corresponding to a specific number of topics for each category and stage. To analyse the results, we reduce the number of topics to 5 by clustering closely related topics using equation \ref{eq:clustering}.

\begin{equation}
T_{c}=\dfrac{\left( \sum ^{N}_{i=1}\sum ^{M}_{j=1}p_{j}x_{j}\right) }{N}
\label{eq:clustering}
\end{equation}

where $N$ refers to the number of topics to be clustered, $M$ represents the number of keywords in each topic, $p_j$ corresponds to the probability of the word $x_i$ in the topic, and $T_c$ is the resultant topic containing the average probabilities of all the words from the $N$ topics. We then represent the top 10 highest probability words in the resultant topic for every category and stage as is shown in Tables \ref{tab:topics_stigma} to \ref{tab:topics_exclude}.

\section{Findings}

Table \ref{tab:topics_stigma}, \ref{tab:topics_offense}, \ref{tab:topics_blame} and \ref{tab:topics_exclude} demonstrate the ten most salient terms related to the generated five topics for each stage (S1, S2, and S3) of four categories, and we summarize each topic through the correlation between the ten terms. We put a question mark for topics from which no pattern can be generated. In general, under the four categories, China and Chinese are always at the centre of discussion. When considering the dynamics across stages, tweets of all four categories extended the discussion to the world situation, and terms representing other nations and races/ethnicities besides China and Chinese started to emerge. 

Notably, the category-based detection and analysis enable us to capture the nuances of themes, and how themes develop through different trajectories across the stages. To specify, the topics in the category of \textit{stigmatization} centre on virus. Discussion tends to associate China and Chinese with the infection and outbreak of virus as well as its negative influences (e.g. emergency; travel). In stage 3, discussion around America became a new focus, with terms trump, president, and propaganda showing up. 

The discussion in the category of \textit{offensiveness} is more political oriented compared to other categories. Especially, in the first two stages, discussion included sensitive political terms concerning China (e.g., hk, uyghyr, taiwan). Besides, ccp (Chinese Communist Party) and human right are two important topics. Only till stage 3, the topics in offensiveness gradually switched the focus to virus. 

The data in the category of \textit{blame} focuses on attributing the cause and consequence of virus to a particular political system (e.g., lie; autocracy, deceit) in the early stages of the discussion. Alike stigmatization, american and president emerged as new topics in stage 3 for the category of blame, although the overall three stages  remained the focus on terms like lie and cover-up by the government. 

The category of exclusion emphasizes virus, trade and human right. Especially, in terms of trade, more negative words are associated with it alongside the development of Covid-19 (e.g. from stop in stage 2 to stop and boycott in stage 3). Additionally, in stage 3, india and indian were related to china under the topic of trade.

\section{Discussion and Conclusions}
Bridging computational methods with social science theories, this research proposes a four-dimensional category for the detection of racist and xenophobic texts in the context of Covid-19. This categorization, combined with a stage wise analysis, enables us to capture the diversity of the topics emerging from racist and xenophobic expression on Twitter, and their dynamic changes across the early stages of Covid-19. This enables the methodological advancement proposed by this research to be transformed into constructive policy suggestions. For instance, as demonstrated in the findings, the topics falling under the category of offensiveness are more likely to be associated with sensitive political issues around China rather than virus in stage 1 and stage 2. Therefore, how to split the discussion of virus from the association of virus with other political topics should draw attention from government of different countries, and this agenda should be incorporated into the official media coverage from the government. Another example is from the category of blame. As shown in the findings, blame usually targets at the transparency of the information from the government (Chinese government especially in early Covid-19). Consequently, it is critical for government of different countries to work on effective and prompt communication with the public under Covid-19. We believe the contribution of this research can be generated beyond the context of Covid-19 to provide insights for future research on racism and xenophobia on digital platforms.  

\bibliographystyle{named}
\bibliography{ijcai21}

\end{document}